\documentclass[10pt,twocolumn,letterpaper]{article}
\usepackage[pagenumbers]{iccv}

\definecolor{iccvblue}{rgb}{0.21,0.49,0.74}
\usepackage[pagebackref,breaklinks,colorlinks,allcolors=iccvblue]{hyperref}

\newcommand{\customfootnotetext}[2]{%
  \begingroup
  \setlength{\skip\footins}{0pt}
  \renewcommand{\thefootnote}{#1}%
  \footnotetext{#2}%
  \endgroup
}

\title{C\textcolor{blue}{$^{3}$}Editor: A\textcolor{blue}{c}hieving \textcolor{blue}{C}ontrollable \textcolor{blue}{C}onsistency in 2D Model for \textcolor{blue}{3}D Editing}

\author{%
  Zeng Tao\textsuperscript{1}\textsuperscript{*} \quad
  Zheng Ding\textsuperscript{2} \quad
  Zeyuan Chen\textsuperscript{2} \quad
  Xiang Zhang\textsuperscript{2} \quad
  Leizhi Li\textsuperscript{2}\quad
  Zhuowen Tu\textsuperscript{2} \\
  \textsuperscript{1}Fudan University \quad 
  \textsuperscript{2}UC San Diego
}

\let\oldtwocolumn\twocolumn
\renewcommand\twocolumn[1][]{%
    \oldtwocolumn[{#1}{
    \begin{center}
           \includegraphics[width=\linewidth,trim=0 2em 0 2em,clip]{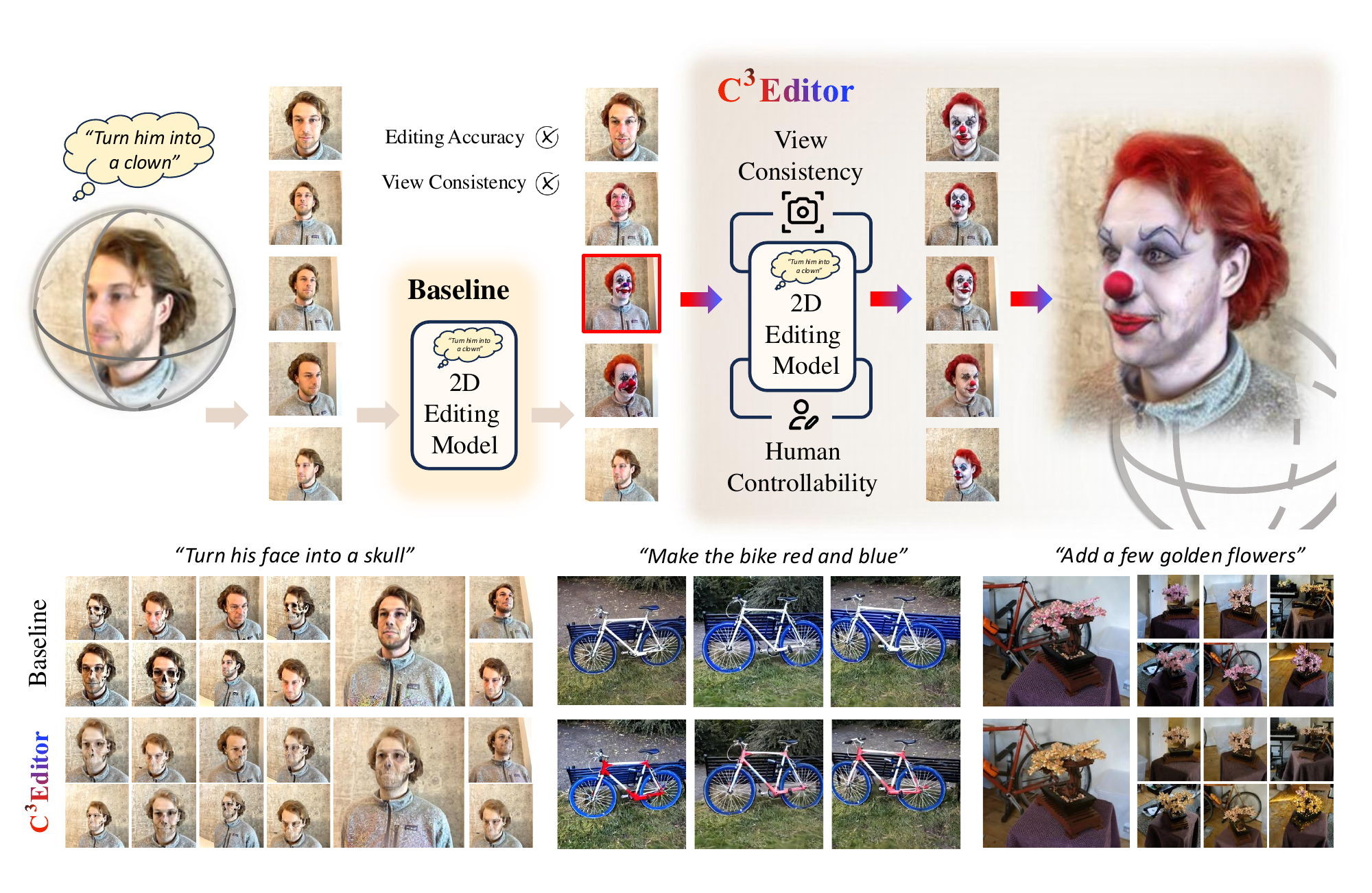}
           \captionof{figure}{\textbf{C$^{3}$Editor: Controllable Consistent 2D Model for 3D Editing}. \textbf{Top:} Our C$^{3}$Editor method generates consistent 2D editing results across different views by following the original 3D scene, editing text, and user guidance, thereby supporting improved 3D editing performance. \textbf{Bottom:} Comparison of 2D and 3D editing results between baseline and C$^{3}$Editor.}
           \label{Fig1}
           \vspace{1em}
        \end{center}
    }]
}

\begin{document}

\maketitle

\customfootnotetext{*}{Work done during internship at UC San Diego.}
\begin{abstract}

Existing 2D-lifting-based 3D editing methods often encounter challenges related to inconsistency, stemming from the lack of view-consistent 2D editing models and the difficulty of ensuring consistent editing across multiple views. To address these issues, we propose C³Editor, a controllable and consistent 2D-lifting-based 3D editing framework. Given an original 3D representation and a text-based editing prompt, our method selectively establishes a view-consistent 2D editing model to achieve superior 3D editing results. The process begins with the controlled selection of a ground truth (GT) view and its corresponding edited image as the optimization target, allowing for user-defined manual edits. Next, we fine-tune the 2D editing model within the GT view and across multiple views to align with the GT-edited image while ensuring multi-view consistency. To meet the distinct requirements of GT view fitting and multi-view consistency, we introduce separate LoRA modules for targeted fine-tuning. Our approach delivers more consistent and controllable 2D and 3D editing results than existing 2D-lifting-based methods, outperforming them in both qualitative and quantitative evaluations.

\end{abstract}

\section{Introduction}

The remarkable success of 2D generative models~\cite{ho2020denoising, kumari2023multi,chefer2023attend,bansal2023universal, avrahami2023chosen,avrahami2023blended,xu2023prompt,zhang2023adding,zhang2024realcompo} has spurred rapid advancements in the field of generation, leading to successful applications in related areas such as editing tasks~\cite{avrahami2022blended,hertz2022prompt,kawar2023imagic,meng2021sdedit,ramesh2022hierarchical,brooks2023instructpix2pix}. Leveraging the superior performance of 2D models as priors also has become a popular approach in 3D tasks~\cite{zhang2023uni,chen2023fantasia3d,lin2023magic3d,metzer2023latent,wang2024prolificdreamer,zhao2025deprdepthguidedsingleview}. Given the scarcity of real-world 3D data and the high cost of training, utilizing pretrained 2D models as guidance offers a promising solution. For example, 2D-lifting-based 3D editing methods~\cite{haque2023instruct} use a 2D editing model~\cite{brooks2023instructpix2pix} to obtain edited images from different viewpoints, which are then used to update the original 3D representation. 

However, directly transferring 2D priors to the 3D domain presents certain challenges, such as the issue of viewpoint consistency~\cite{poole2022dreamfusion}. Since 2D models lack view information and 3D awareness, conflicts between views may arise when applied to 3D tasks. In 3D editing tasks, directly using edited 2D images that lack consistency across views can lead to errors in 3D editing. Some approaches attempt to address this by constructing external datasets~\cite{liu2023zero, shi2023mvdream, liu2024one}. However, addressing the view inconsistency problem in 3D editing remains challenging. The training process requires datasets containing consistent editing text, original 2D images, and edited 2D images across multiple views, which are difficult to obtain. One viable approach is to designate the edited result of one specific view as the ground truth (GT) and, leveraging the generalization ability of the 2D model, gradually adapt other viewpoints to match this viewpoint, achieving internal consistency across views.

Additionally, text-based editing inherently supports diversity, but current 2D-lifting-based 3D editing methods suppress this diversity by uniformly processing different 2D editing results~\cite{haque2023instruct, dong2024vica, chen2024gaussianeditor, chen2024consistdreamer}. Our goal is to allow for the controllability of optimization directions, enabling the 3D editing results to express more possibilities and better align with human intent.

In response, we propose C$^{3}$Editor, a controllable and consistent 2D-lifting-based 3D editing method. Given an original 3D scene and an editing text prompt, we aim to obtain a view-consistent 2D editing model selectively, thereby achieving improved 3D editing results. By selecting a GT view and its corresponding edited image as the optimization target, our approach stabilizes the GT view's editing results and then progressively enforces consistency across different views through view propagation. Furthermore, we introduce separate LoRA modules to fine-tune the model, addressing the unique requirements of GT view fitting and multi-view consistency separately. This structured approach ensures that the 2D editing model achieves cohesive 3D editing results across all views, enhancing both visual consistency and user controllability.

In summary, our contributions are as follows:

\begin{itemize}
    \item We develop a view-consistent 2D editing model based on the original 3D representation and an editing text prompt, facilitating enhanced 3D editing outcomes. This approach effectively bridges the gap between 2D and 3D, as well as between original and edited representations.
    \item Our controllable 3D editing method allows users to select a ground truth (GT) edited image and manually adjust it to produce consistent 3D editing results across views.
    \item Our C$^{3}$Editor method mainly focuses on two aspects: Intra-GT and Inter-view. We specifically design GT selection and intra-GT Loss methods to ensure stable GT fitting, followed by view propagation and inter-view loss for view consistency. Different LoRAs serve separate consistent purposes. Qualitative and quantitative experiments demonstrate the effectiveness of C$^{3}$Editor.
\end{itemize}

\section{Related Work}

\subsection{Diffusion Model and Fine-tuning}

Diffusion models~\cite{ho2020denoising,rombach2022high} have become powerful tools in generative tasks due to their unique approach of iteratively refining data from noise, allowing for precise control over the generation process. These models learn data distributions through a diffusion process that gradually adds and then reverses noise, effectively modeling complex data patterns in images, audio, and even text. Because of their robust performance, diffusion models are widely applied in tasks~\cite{chen2023fantasia3d,lin2023magic3d, metzer2023latent,wang2024prolificdreamer,xu2024bayesian,srivastava2025lay,zeng2025yolo} such as image synthesis, inpainting, super-resolution, and conditional generation, where they can generate or manipulate visual content based on additional inputs, such as text prompts, segmentation maps, or depth maps. This versatility makes them particularly valuable for tasks requiring high-quality, detailed outputs and subtle adjustments.

Fine-tuning diffusion models is essential for adapting them to specific tasks or datasets. Through targeted fine-tuning, diffusion models can be optimized to perform controlled edits, match stylistic demands, or generalize to new domains beyond their original training data. Techniques such as low-rank adaptation (LoRA)~\cite{hu2021lora} and other parameter-efficient tuning methods~\cite{houlsby2019parameter,li2021prefix,liu2021p} allow for effective customization by focusing on updating key parts of the model while keeping the core structure intact. This approach is especially useful when integrating diffusion models as priors in cross-domain applications, where maintaining high fidelity across varying views is critical. Fine-tuning thus enables diffusion models to meet specialized generative requirements, ensuring they maintain both visual quality and flexibility across diverse tasks.

\subsection{Diffusion-based 2D Editing}

Diffusion-based 2D editing techniques~\cite{avrahami2022blended,hertz2022prompt,kawar2023imagic, meng2021sdedit,ramesh2022hierarchical,brooks2023instructpix2pix} have revolutionized the field of image manipulation by leveraging the denoising diffusion process to transform noise into structured visual representations. In these models, editing is performed iteratively, where each step refines the image by reversing the noise and generating realistic features, allowing for adequate control over the level and type of modifications applied.

The key advantage of diffusion-based 2D editing lies in its ability to use conditional inputs, like text prompts or segmentation maps, to guide the editing process. For example, Instruct-Pix2Pix~\cite{brooks2023instructpix2pix} can interpret prompts to modify colors, add textures, or alter structures while maintaining the coherence of the image. These models can learn data distributions that align with specific editing goals, making them versatile across diverse applications. By fine-tuning or adjusting model parameters, diffusion models can also be specialized for specific editing tasks, allowing them to adapt to particular styles or constraints required by the user. This combination of iterative refinement, conditional control, and adaptability has made diffusion-based 2D editing a powerful tool in modern image generation and editing tasks.

\subsection{2D-lifting-based 3D Editing}

Recent advancements in 3D editing have increasingly integrated diffusion-based 2D editing models, leveraging their established capabilities to enhance 3D workflows~\cite{cao2024mvinpainter,chen2024proedit, chen2024gaussianeditor,haque2023instruct,dong2024vica,chen2024consistdreamer}. These models, originally designed for detailed image modifications, contribute to 3D editing by transferring their proficiency in nuanced, high-quality adjustments to three-dimensional representations. Methods like Neural Radiance Fields (NeRF)~\cite{mildenhall2021nerf} and 3D Gaussian Splatting (3DGS)~\cite{kerbl20233d} incorporate 2D editing models to improve the consistency and detail of 3D content.

By using 2D diffusion models as priors, recent approaches enhance the fidelity and stylistic consistency of 3D edits, especially in maintaining coherence across multiple views. Some works, such as DGE~\cite{chen2024dge}, combine images from different viewpoints into videos for processing. A primary challenge in this domain remains ensuring multi-view consistency, as traditional 2D-based edits applied to 3D models often lead to discrepancies between perspectives. Some methods, such as ConsistDreamer~\cite{chen2024consistdreamer}, model 3D-aware consistency by means of constraints like neural feature alignment or volume-based feature consistency. This has provided inspiration for our work. However, since it is not open-sourced, it is impossible to make a comparison for now. We compare our method with NeRF-based Instruct-NeRF-to-NeRF~\cite{haque2023instruct}, ViCA-NeRF~\cite{dong2024vica}, and GS-based GaussianEditor~\cite{chen2024gaussianeditor}.

\begin{figure*}[t]
    \centering
    \includegraphics[width=\linewidth,trim=1em 1em 1em 1em,clip]{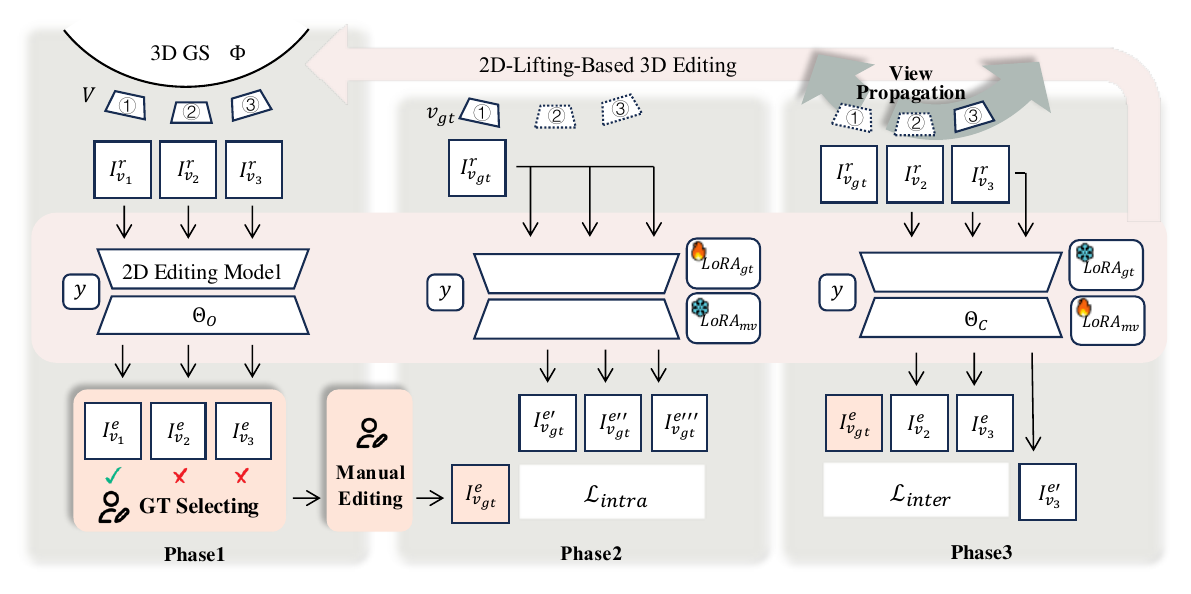}
    \caption{\textbf{C$^{3}$Editor Method Pipeline}. Given a 3D representation $\Phi$, a text prompt for editing $y$, and the original 2D editing model $\Theta_{O}$, our method aims to process $\Theta_{O}$ to obtain $\Theta_{C}$ that is related to $y$ and ensures multi-view consistency, thereby achieving improved 3D editing results. \textbf{Phase 1}: Controllable optimization direction selecting and manual editing in \cref{sec:optimizationdirection}. \textbf{Phase 2}: Intra-GT prior fitting in \cref{sec:priorfitting} to fit the GT information. \textbf{Phase 3}: View propagation and inter-view consistent construcing in \cref{sec:viewpropagation}. Details of LoRA modules for separate fine-tuning are in \cref{sec:paralleltuning}.}
    \label{fig:method}
\end{figure*}

\section{Method}
\label{sec:method}

\subsection{Overview}

Given a 3D representation $\Phi$ (\eg, 3D GS), a text prompt for editing $y$, and the original 2D editing model $\Theta_{O}$ (like Instruct-Pix2Pix~\cite{brooks2023instructpix2pix}), the goal of our method is to process $\Theta_{O}$ to obtain $\Theta_{C}$ that is related to $y$ and ensures multi-view consistency, thereby achieving improved 3D editing results. In \cref{sec:optimizationdirection}, the ground truth (GT) view $v_\mathrm{gt}$ and GT edited image $I^{e}_{v_\mathrm{gt}}$ are manually selected from the 2D editing results $I^{e}_{v}$ of different views $v$, rendered by $\Theta_{O}$, which serve as the optimization target. In \cref{sec:priorfitting}, we optimize $\Theta_{O}$ with a specifically designed intra-GT loss to fit $I^{e}_{v_\mathrm{gt}}$. We maintain global consistency through the view propagation method and inter-view loss described in \cref{sec:viewpropagation}. In \cref{sec:paralleltuning}, we introduce different LoRAs for different fine-tuning objectives to separately fine-tune the diffusion model.

Using $\Theta_{C}$ obtained, each view $v$ in $V$ undergoes a complete editing process, producing consistent view-editing results $I^{e}_{v}$. Considering the gradient storage issues in fine-tuning the diffusion model, each complete editing process includes 5 diffusion denoising steps, which achieves a good trade-off between GPU memory limits and editing quality. The final 3D editing result is obtained by updating the original 3D representation $\Phi$ with the edited results of all views $I^{e}_{v}$. We adopt 3D Gaussian Splatting (3D GS) as our 3D representation due to its efficient training speed and excellent rendering quality. We adopt widely used Instruct-Pix2Pix~\cite{brooks2023instructpix2pix} as our diffusion-based pre-trained 2D editing model, for its outstanding performance in 2D editing tasks. The method for updating 3D GS is consistent with that in GaussianEditor. The detailed process is illustrated in \cref{fig:method}.

\subsection{Controllable Optimization Direction}
\label{sec:optimizationdirection}

The independent 2D editing processes of different views $v$ lead to different editing results. To avoid view conflicts in 3D editing, we select the editing result $I^{e}_{v_\mathrm{gt}}$ from a specific view $v_\mathrm{gt}$ as the optimization direction. In subsequent operations, the 2D editing model will use this GT as a reference to edit images from other views, thereby preventing conflicts in the 3D editing process.

As shown in \cref{fig:method} Phase 1, for each view $v$, an independent editing process is performed, resulting in different editing outcomes $I^{e}_{v}$. User then selects a specific view and its corresponding edited result as the GT view $v_\mathrm{gt}$ and GT edited image $I^{e}_{v_\mathrm{gt}}$, setting the target optimization direction. Different choices of view and edited results lead to different optimization directions, and consequently, varying final 3D editing outcomes, which are shown in \cref{sec:controllableediting}. Therefore, this selection should follow certain guidelines, such as choosing results of higher editing quality and selecting a more central view. Based on these guidelines, the user can choose their desired optimization direction.

For the obtained GT image $I^{e}_{v_\mathrm{gt}}$, users can directly use it as is. However, if there are any unsatisfactory elements, users can make manual edits according to their preferences. They can utilize image editing tools such as Photoshop to modify the image content, then set the edited image as the GT image $I^{e}_{v_\mathrm{gt}}$ for the model. The GT view and edited result are then used to guide the subsequent 2D editing model fine-tuning process, ensuring that the 2D model can achieve controllable editing results across all views.

\subsection{Intra-GT Prior Fitting}
\label{sec:priorfitting}

With the optimization direction established, the next step is to train the 2D diffusion model to fit the GT image. Adjustments to the 2D diffusion model are divided into two parts: intra-GT and inter-view adaptation. In this section, we need to make the 2D model fit the chosen optimization direction $I^{e}_{v_\mathrm{gt}}$ on the GT view $v_\mathrm{gt}$, aiming to establish a foundational intra-view editing stability on the GT view $v_\mathrm{gt}$.

As shown in \cref{fig:method} Phase 2, we freeze the 3D representation and add LoRA modules to the diffusion model for fine-tuning. The edited image $I^{e}_{v_\mathrm{gt}}$ is used as the GT. An independent, complete editing process is performed on the rendered image $I^{r}_{v_\mathrm{gt}}$ of $v_\mathrm{gt}$ to obtain an edited image $I^{e'}_{v_\mathrm{gt}}$ that differs from the GT image $I^{e}_{v_\mathrm{gt}}$. Compute the loss between $I^{e'}_{v_\mathrm{gt}}$ and $I^{e}_{v_\mathrm{gt}}$, back-propagate, and update the LoRA. The loss $\mathcal{L}_\mathrm{intra}$ consists of two parts: the $L_1$ loss and perceptual loss between $I^{e'}_{v_\mathrm{gt}}$ and $I^{e}_{v_\mathrm{gt}}$.

\begin{equation}
    \mathcal{L}_\mathrm{intra} = \lambda_{1} L_{1}(I^{e'}_{v_\mathrm{gt}}, I^{e}_{v_\mathrm{gt}}) + \lambda_{2}  L_\mathrm{Perceptual}(I^{e'}_{v_\mathrm{gt}}, I^{e}_{v_\mathrm{gt}})
\end{equation}

Through multiple iterations of this process, the fine-tuned 2D diffusion model acquires a certain fitting capability for the GT image, while also improving editing stability for the same view, achieving similar results across different editing processes.

\subsection{View Propagation and Inter-view Consistency}
\label{sec:viewpropagation}

After \cref{sec:priorfitting}, the 2D editing model can only fit $I^{e}_{v_\mathrm{gt}}$ on $v_\mathrm{gt}$, with limited generalization ability, and has limited consistency in editing effects for views that differ significantly. If the current 2D model is used directly as the prior, it can only maintain consistent editing for $v_\mathrm{gt}$ and views nearby, while its performance on more distant views remains uncertain. Therefore, in this section, we introduce additional methods to ensure consistent editing across all views. We leverage the interrelations between viewpoints to enable the 2D diffusion model to achieve consistent editing across all views gradually.

\begin{figure*}[t]
    \centering
    \includegraphics[width=\linewidth]{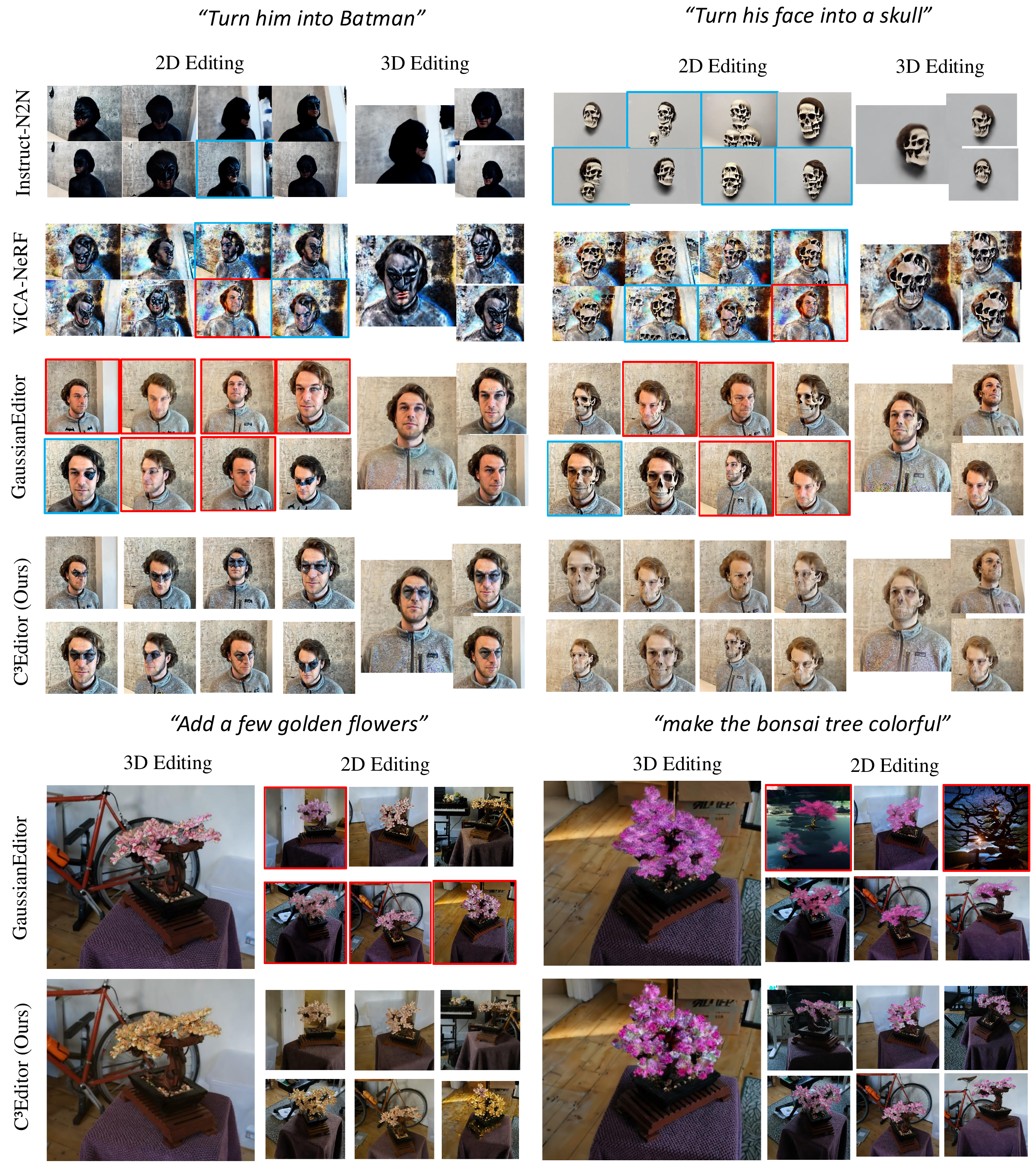}
    \caption{\textbf{Comparison of Qualitative Results}. Compared to baseline methods, C$^{3}$Editor can generate view-consistent 2D images, avoiding inter-view conflicts (highlighted in \textcolor{blue}{blue}) and erroneous 2D edits (highlighted in \textcolor{red}{red}), thereby achieving better 3D editing results.}
    \label{fig:qualitative}
\end{figure*}

As shown in \cref{fig:method} Phase 3, specifically, we sort the views as a sequence $S$ by their distance of camera center points from $v_\mathrm{gt}$, from closest to farthest, and then perform fine-tuning of the 2D diffusion model on each view in $S$ other than $v_\mathrm{gt}$. $v_{0}$ in the sequence represents $v_\mathrm{gt}$. We perform a 2D editing process on each view $v_{i}$ with the index $i\in \{1,2,\dots,j,i,\dots,n-1\}$ in the sequence separately. The resulting image $I^{e}_{v_{i}}$ serves as the GT image for $v_{i}$. Next, an independent 2D editing process is applied to this view, and another edited image $I^{e'}_{v_{i}}$ is obtained. The loss $\mathcal{L}_\mathrm{inter}$ comprises three parts: $\mathrm{loss}\,1$ between the edited image $I^{e'}_{v_{i}}$ and the GT image $I^{e}_{v_{i}}$, $\mathrm{loss}\,2$ between $I^{e'}_{v_{i}}$ and $I^{e}_{v_{j}}$ of the closest processed view $v_{j}$, $\mathrm{loss}\,3$ between $I^{e'}_{v_{i}}$ and $I^{e}_{v_\mathrm{gt}}$. $\mathcal{L}_\mathrm{inter}$ is as follows:
\begin{align}
    \mathcal{L}_\mathrm{inter} = & \underbrace{\lambda_{3}  L_1(I^{e'}_{v_{i}}, I^{e}_{v_{i}}) + \lambda_{4} L_\mathrm{Perceptual}(I^{e'}_{v_{i}}, I^{e}_{v_{i}})}_{\mathrm{loss}\,1} \nonumber \\
    & + \underbrace{\lambda_{5} L_\mathrm{Perceptual}(I^{e'}_{v_{i}}, I^{e}_{v_{j}})}_{\mathrm{loss}\,2} \nonumber \\
    & + \underbrace{\lambda_{6} L_\mathrm{Perceptual}(I^{e'}_{v_{i}}, I^{e}_{v_\mathrm{gt}})}_{\mathrm{loss}\,3}
\end{align}
$\mathcal{L}_\mathrm{inter}$ is back-propagated, and LoRA is used to fine-tune the diffusion model. After this process, we reverse the sequence $S$ and repeat the above steps until reaching $v_\mathrm{gt}$. The generalization capability gradually expands from the $v_\mathrm{gt}$ to encompass all views.

\subsection{Separate Fine-tuning}
\label{sec:paralleltuning}

To prevent the loss of GT information during the inter-view fine-tuning process, we design two LoRAs, each serving different fine-tuning goals. The fine-tuning of the 2D Editing model $\Theta_{O}$ is divided into two main aspects: LoRA$_\mathrm{gt}$ for fitting the GT view image $I^{e}_{v_\mathrm{gt}}$, and LoRA$_\mathrm{mv}$ for ensuring consistency across different views. The inference process takes place in \cref{sec:optimizationdirection} and \cref{fig:method} Phase 1, with no trainable model parameters. In \cref{sec:priorfitting} and \cref{fig:method} Phase 2, we use LoRA$_\mathrm{gt}$ to fine-tune $\Theta_{O}$ while keeping LoRA$_\mathrm{mv}$ frozen. After this step, LoRA$_\mathrm{gt}$ helps the $\Theta_{O}$ fit $I^{e}_{v_\mathrm{gt}}$. In \cref{sec:viewpropagation} and \cref{fig:method} Phase 3, we freeze LoRA$_\mathrm{gt}$ and use LoRA$_\mathrm{mv}$ to fine-tune $\Theta_{O}$. During the separate fine-tuning process, the model uses the GT information obtained by LoRA$_\mathrm{gt}$ and leverages LoRA$_\mathrm{mv}$ to achieve global consistency.

\section{Experiments}
\subsection{Implementation Details}

Our method builds on the advanced 2D-lifting-based 3D GS Editing Method, GaussianEditor~\cite{chen2024gaussianeditor}. Specifically, we use 3D GS~\cite{kerbl20233d} as the 3D representation and the widely-used Instruct-Pix2Pix~\cite{brooks2023instructpix2pix} as the diffusion-based 2D editing model. All experiments were conducted on a single NVIDIA RTX A6000, with the fine-tuning process taking 1 minute in total. We use MipNeRF-360~\cite{barron2022mip} and Instruct-NeRF-to-NeRF dataset~\cite{haque2023instruct} to measure the performance of our method. The MipNeRF-360 dataset contains 360-degree views of 3D scenes, while the Instruct-NeRF-to-NeRF dataset contains 3D scenes. We use the CLIP-Score~\cite{taited2023CLIPScore} (image-text and image-image) as the evaluation metrics. The former measures the similarity between 3D edited results and editing text, while the latter measures the similarity between 2D images produced by the 2D editing process. A higher score indicates greater editing quality and view consistency. We also use the Fréchet Inception Distance (FID)~\cite{heusel2017gans, Seitzer2020FID} between original rendered images and edited results to evaluate the quality of 3D editing. A lower FID score indicates higher image quality.

For each scene and editing text, we perform unique training to obtain the corresponding 2D editing model. Following the same approach as GaussianEditor, we first use Gaussian Semantic Tracing~\cite{chen2024gaussianeditor} to generate a mask of the editing target within the 3D GS. We then follow the processing steps outlined in \cref{sec:method} to obtain a 2D editing model with view consistency. This model serves as the 2D prior, achieving the final 3D editing result. For further details on the 3D editing process, please refer to GaussianEditor~\cite{chen2024gaussianeditor}.

The hyperparameters are set as follows: 30 iterations for $\mathcal{L}_\mathrm{intra}$ updates, and 3 iterations for $\mathcal{L}_\mathrm{inter}$ updates. We set $\lambda_{n}$=1. For the two LoRA modules, we use the AdamW optimizer with the following settings: $r=4$, $\text{LoRA alpha}=4$, init LoRA weights="gaussian", $\text{lr}=10^{-4}$, $\text{betas}=(0.9, 0.999)$, $\text{weight decay}=10^{-2}$, $\text{eps}=10^{-8}$.

\subsection{Comparison with Baselines}

\noindent \textbf{Qualitative Comparison}. \cref{fig:qualitative} illustrates the qualitative results of our method. Compared to NeRF-based Instruct-NeRF-to-NeRF~\cite{haque2023instruct}, ViCA-NeRF~\cite{dong2024vica}, and GS-based GaussianEditor~\cite{chen2024gaussianeditor} methods, our approach demonstrates superior consistency in both 2D and 3D editing. With the editing texts of the face scene, such as ``Turn him into Batman" and ``Turn his face into a skull", our method addresses inconsistency issues encountered by baseline methods, achieving accurate edits on facial features. In natural scenes, like the bonsai scene, our approach produces improved editing results compared to baseline methods, excelling in color and other details. This improvement is due to the fact that current 3D editing techniques using 2D editing models as priors often encounter inconsistencies during 2D editing, leading to issues such as inter-view inconsistency (highlighted in the \textcolor{blue}{blue} boxes) and editing errors (highlighted in the \textcolor{red}{red} boxes), which result in inaccurate or incomplete 3D edits. Our method, however, achieves consistent 2D editing results, leading to accurate 3D edits. For more qualitative results, please refer to the following sections and supplementary materials.

\begin{table}[h]
    \centering
    \resizebox{\linewidth}{!}{
    \begin{tabular}{c|c|c}
        \toprule
        Method & GaussianEditor & \textbf{C$^{3}$Editor (Ours)} \\
        \midrule
        Image-Text CLIP-Score ($\uparrow$) & 24.18 & \textbf{25.21} \\
        \midrule
        Image-Image CLIP-Score ($\uparrow$) & 84.20 &  \textbf{87.46} \\
        \midrule
        FID ($\downarrow$) & 112.21 & \textbf{89.95} \\
        \midrule \midrule
        Time Difference & \multicolumn{2}{|c}{Avg 56s more than GaussianEditor}  \\
        \bottomrule
    \end{tabular}}
    \caption{\textbf{Comparison of Quantitative Results}. Our method surpasses the baseline method on all of the three metrics.}
    \label{tab:quantitative}
\end{table}

\noindent \textbf{Quantitative Comparison}. As shown in \cref{tab:quantitative}, we use CLIP-Score~\cite{taited2023CLIPScore} (image-image and image-text) and FID~\cite{heusel2017gans, Seitzer2020FID} as the metrics for quantitative evaluation. Specifically, we calculate the CLIP-Score between images from edited 3D results and editing text. A higher score indicates better editing loyalty. Our method achieves a higher CLIP-Score, demonstrating the improved quality of our 3D editing results. We also calculate the CLIP-Score within the 2D images produced by the editing process. A higher score indicates greater similarity between edited 2D images, thus representing stronger view consistency. Our method achieves a higher CLIP-Score, demonstrating the view consistency of our editing approach. The lower FID score of our method indicates better image quality in the 3D editing results. Our method outperforms GaussianEditor in both qualitative and quantitative evaluations, showcasing the effectiveness of our approach in achieving controllable and consistent 3D editing results.

\subsection{Controllable Editing}
\label{sec:controllableediting}

\begin{figure}[h]
    \centering
    \includegraphics[width=\linewidth]{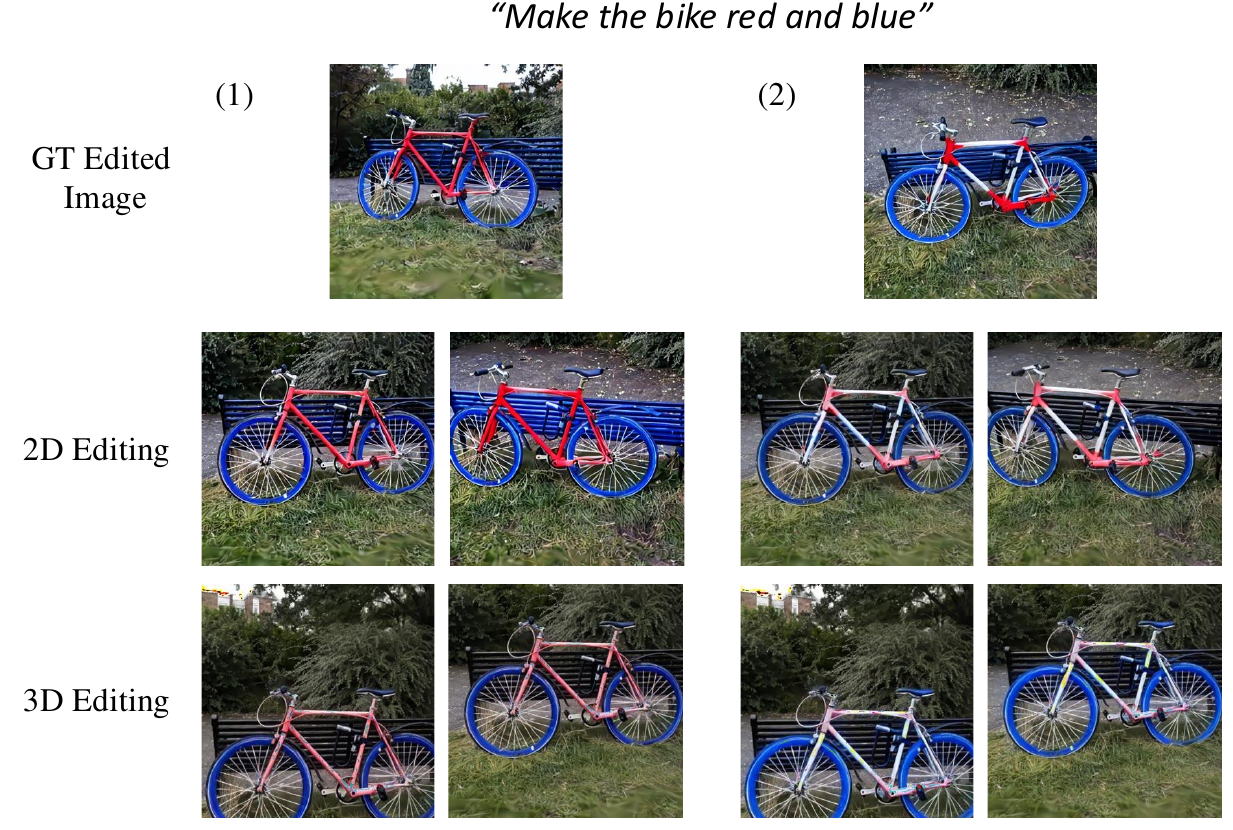}
    \caption{\textbf{Controllable Editing Results with Different GT Selections}. In C$^{3}$Editor, users can decide the optimization direction by selecting the GT edited image they prefer.}
    \label{fig:gt}
\end{figure}

\noindent \textbf{Controllable GT Selection}. With different selections of the GT edited image, our method can achieve the corresponding editing results. As shown in \cref{fig:gt}, given the editing prompt ``Make the bike red and blue", the pre-trained 2D editing model can produce different outcomes. In \cref{fig:gt} (1), the wheels are edited to blue while the entire bike frame is edited to red. In \cref{fig:gt} (2), the wheels are also blue, but only part of the frame is edited to red. Using our method, the obtained 2D editing model can edit images from other views to produce the corresponding 3D editing result. 

The 2D editing process inherently allows for diverse outcomes, and our controllable GT selection effectively supports this diversity, enabling results that better align with user intentions. By allowing targeted selection of the GT image, our approach minimizes 3D editing errors that may arise from inaccuracies in 2D editing, thus enhancing the precision and stability of the final editing outcome.

\begin{figure}[h]
    \centering
    \includegraphics[width=\linewidth]{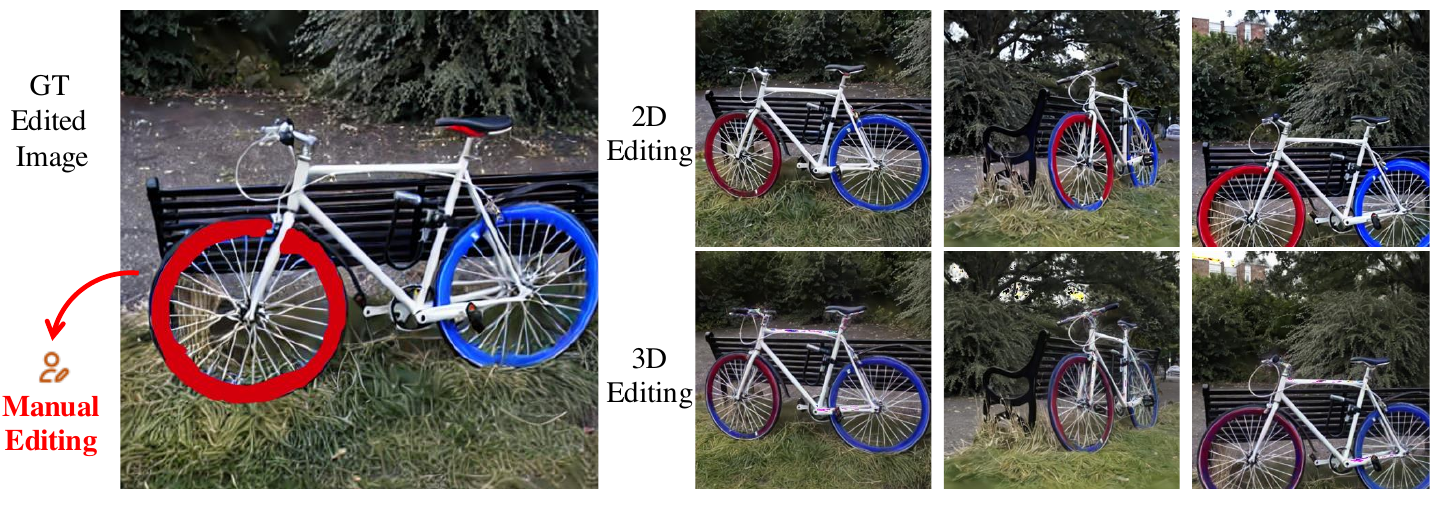}
    \caption{\textbf{Controllable Editing Results with Manual Editing}. In C$^{3}$Editor, users can edit the GT manually and obtain the corresponding 2D and 3D editing results.}
    \label{fig:manual}
\end{figure}

\noindent \textbf{Manual GT Editing}. Furthermore, users can manually edit the GT image according to their preferences. As shown in \cref{fig:manual}, for the editing prompt ``Make the bike red and blue", we modify the original edit to turn the front wheel of the bike red. Using this manually edited image as the GT image $I^{e}_{v_\mathrm{gt}}$, we proceed with the subsequent steps. The final 2D and 3D editing results maintain consistency with the GT image and exhibit view consistency. Our method offers users a manual editing option, enabling them to correct 2D editing results and align the model’s output with human intent. This feature introduces an additional dimension of controllable generation, allowing for enhanced customization and adaptability in the editing process.

\subsection{Ablation Study}

\begin{figure}[h]
    \centering
    \includegraphics[width=\linewidth]{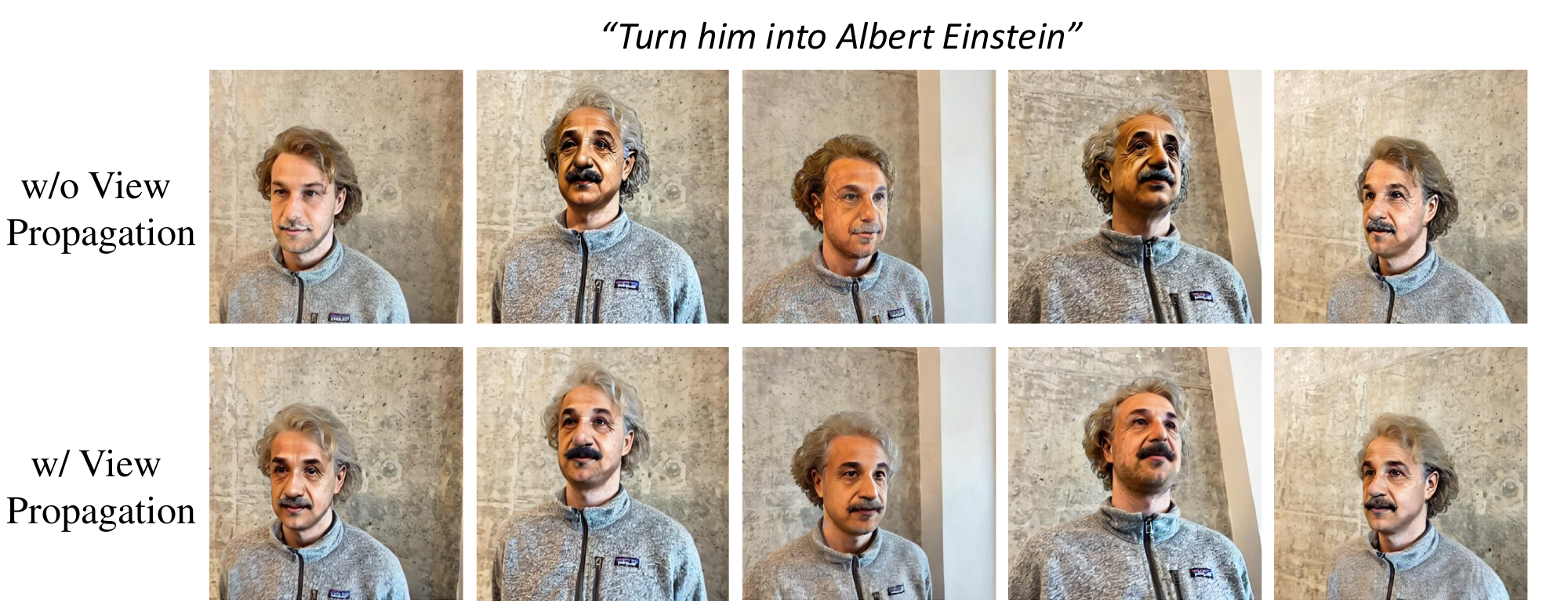}
    \caption{\textbf{Ablation Study on View Propagation}. View propagation helps obtain more view-consistent results than the GT view.}
    \label{fig:random}
\end{figure}

\noindent \textbf{Effectiveness of View Propagation}. We conduct ablation experiments to evaluate the effectiveness of view propagation. As shown in \cref{fig:random}, after fine-tuning the 2D diffusion model based on the GT image $I^{e}_{v_\mathrm{gt}}$, we sort one set of views in order based on their camera center distance to the GT view $v_\mathrm{gt}$ and another set in random order. We then proceed with subsequent steps under these different viewpoint orders. It can be observed that without view propagation, the 2D diffusion model could not achieve fully consistent results. However, with view propagation, the consistency significantly improved. This is because, once the model is fitted to the GT, it gains the ability to produce stable outputs for the GT. It also exhibits a certain level of generalization for viewpoints close to the GT camera position. However, for more distant viewpoints, due to the large gap between input images $I^{r}_{v}$, it is unable to achieve a consistent editing result.

\begin{table}[h]
    \centering
    \resizebox{\linewidth}{!}{
    \begin{tabular}{c|c|c}
        \toprule
        Method & LoRA & LoRA$_\mathrm{gt}$ + LoRA$_\mathrm{mv}$ \\
        \midrule
        Image-Image CLIP-Score ($\uparrow$) & 87.03 &  \textbf{87.46} \\
        \bottomrule
    \end{tabular}}
    \caption{\textbf{Ablation Study on Separate Fine-Tuning}. Using different LoRAs to separately fine-tune the diffusion model can achieve better performance on view consistency.}
    \label{tab:lora}
\end{table}

\noindent \textbf{Effectiveness of Separate LoRA Fine-tuning}. We also conduct ablation on the design of LoRA. As shown in \cref{tab:lora}, we compare results obtained using only a single LoRA with those achieved using different LoRAs for fine-tuning different parts. We use the image-to-image CLIP-Score as the evaluation metric. It is observed that when using two LoRAs for fine-tuning different components, our method produces better results. This is because fine-tuning the diffusion model on the same LoRA can cause disturbances to the previously acquired GT information during subsequent viewpoint fine-tuning, thereby reducing inter-view editing consistency.

\section{Visualization}
\begin{figure}[h]
    \centering
    \includegraphics[width=\linewidth]{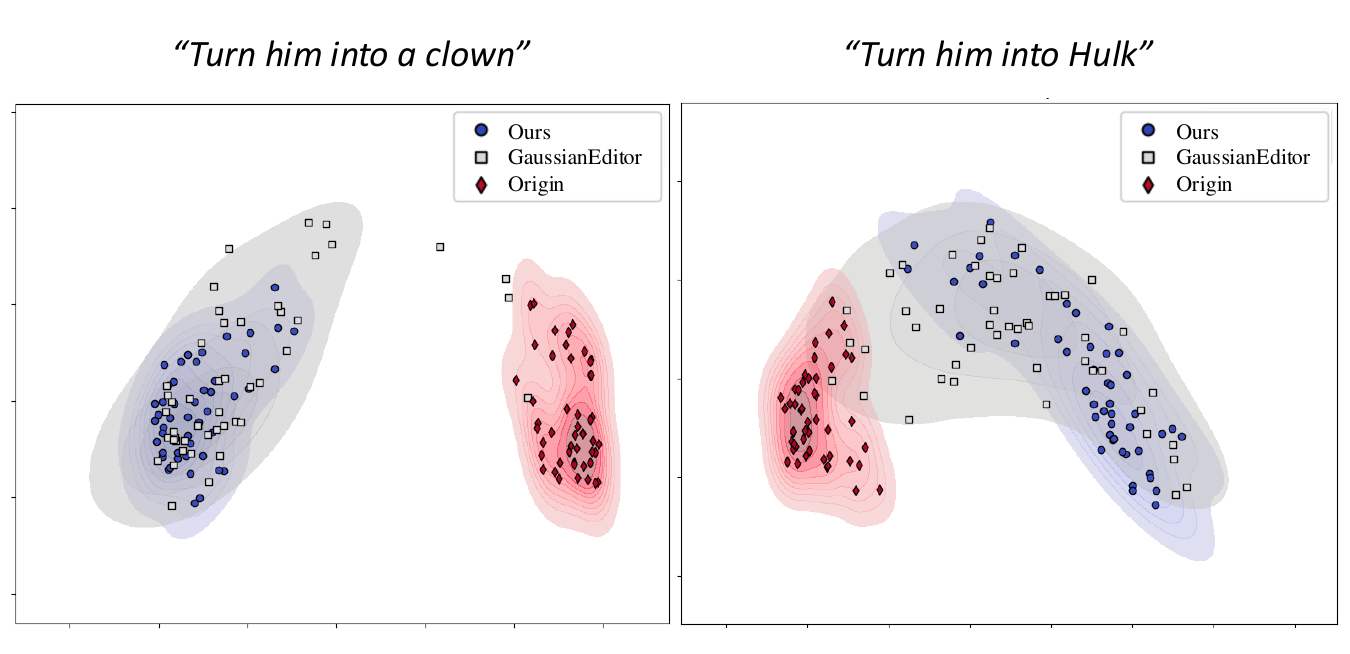}
    \caption{\textbf{Visualization of Original and Edited Image Features}. Features of edited images obtained by C$^{3}$Editor are more concentrated than the baseline model.}
    \label{fig:visualization}
\end{figure}

\noindent \textbf{Visualization of Rendered and Edited Image Features}. We visualized the image features resulting from our edits in \cref{fig:visualization}. Each point in the figure represents an image after feature extraction and dimensionality reduction. Each 2D image was feature-extracted using CLIP ViT-B/32~\cite{radford2021learning}, followed by PCA for dimensionality reduction, and these features were plotted as 2D scatter plots and density plots. In the figures, blue represents images generated by our method, gray represents those generated by the baseline method, GaussianEditor, and red represents the original rendered images. The editing prompt on the left is ``Turn him into a clown," while on the right, it is ``Turn him into Hulk." As shown, the features of the 2D edited images produced by our method are more concentrated than those from the baseline method, indicating that our method achieves stronger view consistency, nearly matching the original images' consistency. Additionally, the features generated by our method show almost no outliers or points that are confused with the original image features, demonstrating that our approach avoids the erroneous edits seen in the baseline method. Our method not only improves the consistency of the generated images but also reduces the occurrence of incorrect edits.

\begin{figure}[h]
    \centering
    \includegraphics[width=\linewidth]{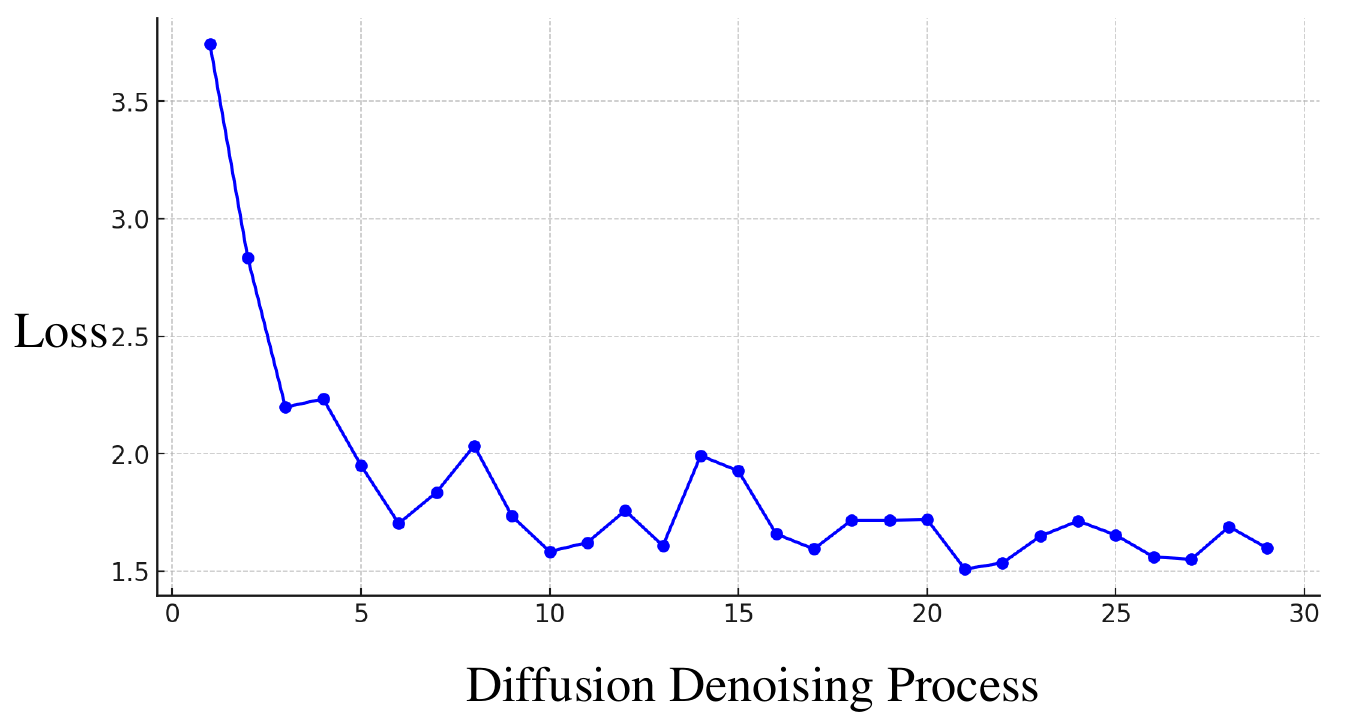}
    \caption{\textbf{Visualization of Loss Change During Intra-GT Prior Fitting}. The loss gradually decreases as the iterations progress, indicating that the fine-tuning process effectively stabilizes the editing results for the GT view.}
    \label{fig:loss}
\end{figure}

\noindent \textbf{Visualization of Intra-GT Prior Fitting}. \cref{fig:loss} illustrates the change in $\mathcal{L}_\mathrm{intra}$ during the Intra-GT Prior Fitting phase. In this process, our goal is for each independently performed diffusion denoising process on the GT view to approximate the GT edited image. As shown in the figure, the loss gradually decreases as the iterations progress. This indicates that the fine-tuning process effectively stabilizes the editing results for the GT view, consistently producing outputs close to the GT edited image. At this phase, the 2D editing model increasingly captures the GT information and achieves stable editing on the GT view.

\section{Conclusion}

In this paper, we propose C$^{3}$Editor, a controllable and consistent 2D-lifting-based 3D editing method. Our approach creates the specific 2D editing model to assist in achieving view consistency and controllable 3D editing results. Qualitative and quantitative evaluations demonstrate that our method outperforms baseline methods in both 2D and 3D results.

\noindent \textbf{Limitations}. Our method still has certain limitations. For instance, a unique 2D editing model must be trained for each specific scene and editing prompt. Meanwhile, the 3D optimization process may also result in the outcome not being fully consistent with the GT edited image. In the future, we aim to enhance the generalization capabilities of the editing model and move towards developing a truly generic multi-view 3D editing model.

\paragraph{Acknowledgment} This work is supported by NSF award IIS-2127544 and NSF award IIS-2433768.

{
    \small
    \bibliographystyle{ieeenat_fullname}
    \bibliography{main}
}

\appendix

\section{Appendix}

\subsection{Overview}

\cref{sec:suppl_implementation} provides additional implementation details of our method C$^3$Editor, as in Sec. 4.1 in \textit{Main Paper}. \cref{sec:suppl_qualitative} provides additional qualitative results of our method C$^3$Editor, as in Sec. 4.2 in \textit{Main Paper}. \cref{sec:suppl_ablation} provides the ablation study of view propagation in our method C$^3$Editor, as in Sec. 4.4 in \textit{Main Paper}. \cref{sec:suppl_gt} provides the visualization of ground truth fitting in our method C$^3$Editor, as in Sec. 5 in \textit{Main Paper}. \cref{sec:suppl_prompt} provides the prompt library used in our method C$^3$Editor, as in Sec. 4.2 in \textit{Main Paper}. The videos in the \textit{Suppl} folder demonstrate our editing demo. The editing prompt is: turn him into a clown, and the selected GT view is Frame 40.

\begin{figure}[htb]
    \centering
    \includegraphics[width=\linewidth,trim=0 1em 0 0,clip]{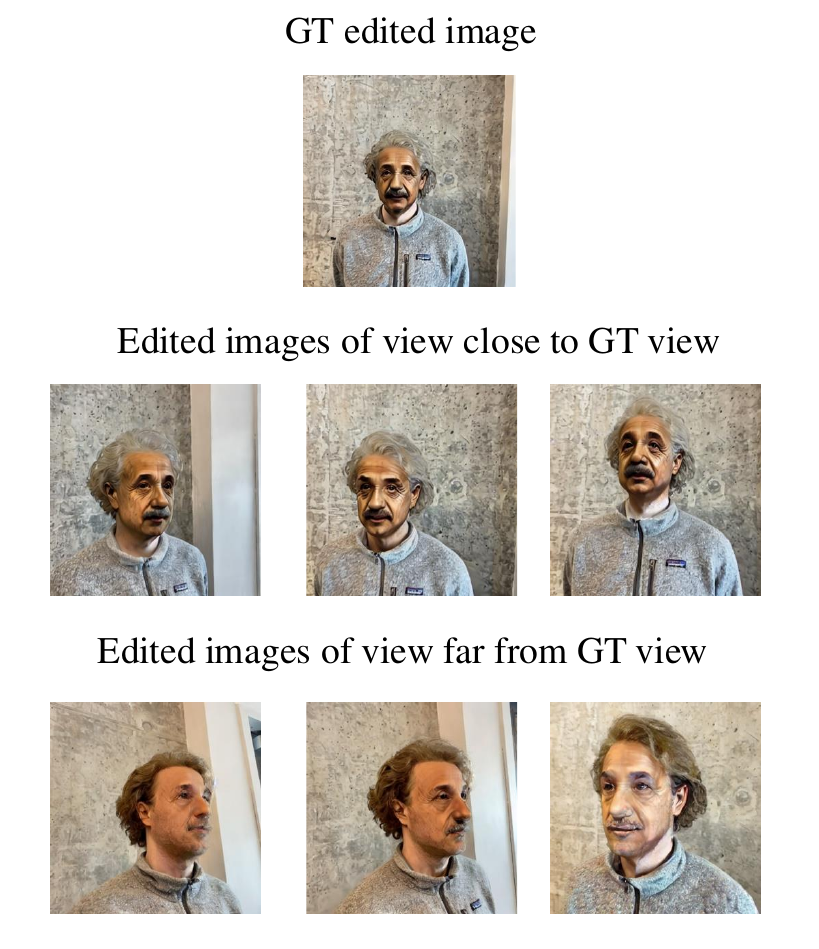}
    \caption{2D editing results without view propagation.}
    \label{fig:suppl_gt_view_prop}
\end{figure}

\begin{figure*}[htb]
    \centering
    \includegraphics[width=\linewidth,trim=0 1em 0 0,clip]{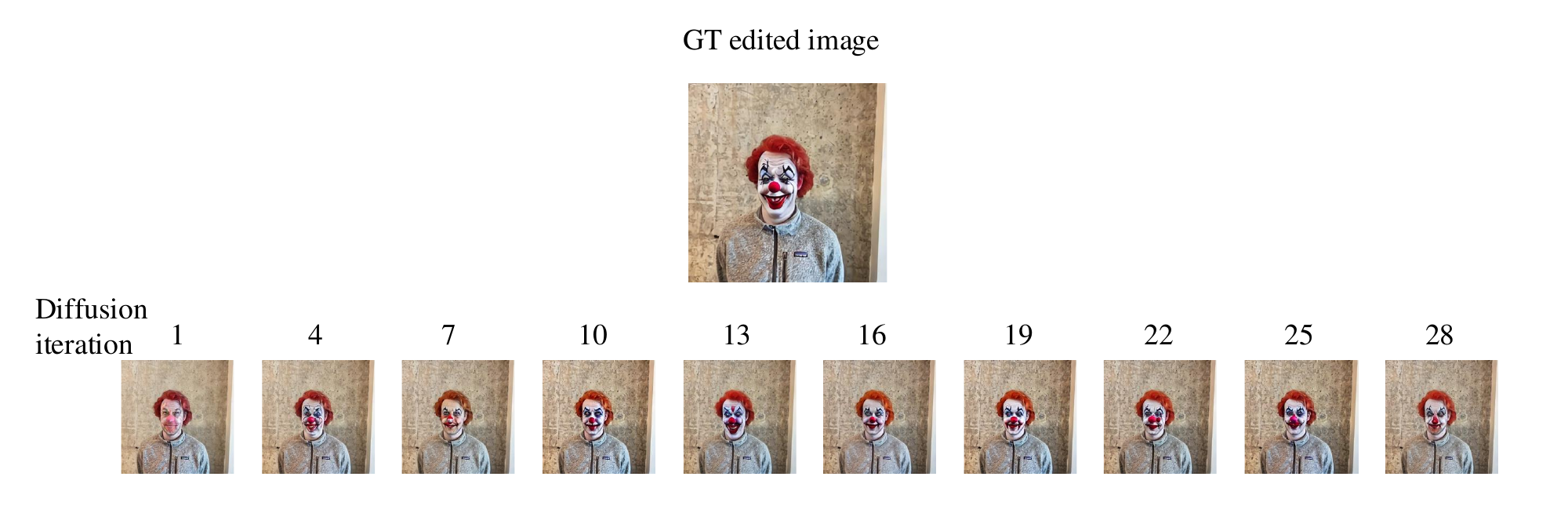}
    \caption{2D editing results of the GT view during intra-GT prior fitting}
    \label{fig:suppl_fitting}
\end{figure*}

\subsection{Implementation Details}
\label{sec:suppl_implementation}

As detailed in Sec. 4.1 of the \textit{Main Paper}, our approach builds upon the advanced 2D-lifting-based 3D Gaussian Splatting (GS) Editing framework, GaussianEditor~\cite{chen2024gaussianeditor}. Specifically, we adopt 3D Gaussian Splatting~\cite{kerbl20233d} as the underlying 3D representation and leverage Instruct-Pix2Pix~\cite{brooks2023instructpix2pix}, a state-of-the-art diffusion-based 2D editing model. All experiments were executed on a single NVIDIA RTX A6000 GPU, with the fine-tuning process requiring only 1 minute in total.

We evaluate our method using the MipNeRF-360~\cite{barron2022mip} and Instruct-NeRF-to-NeRF~\cite{haque2023instruct} datasets. The MipNeRF-360 dataset provides 360-degree views of various 3D scenes, while the Instruct-NeRF-to-NeRF dataset consists of diverse 3D editing scenarios. To assess performance, we utilize CLIP-Scores~\cite{taited2023CLIPScore} (both image-text and image-image) and Fréchet Inception Distance (FID)~\cite{heusel2017gans, Seitzer2020FID}. CLIP-Score (image-text) evaluates the alignment between 3D edited outputs and the provided editing text, while CLIP-Score (image-image) measures consistency across 2D views generated during the editing process. Higher CLIP-Scores indicate better editing quality and greater view consistency. FID, computed between the original rendered images and the edited results, serves as a metric for image quality, where lower scores signify superior results. For the image-image CLIP-score across $n$ views, we compute the similarity between view $i$ and view $i+1$ and take the average as the score for the given setting (view $n$ is computed with view $0$). For the image-text CLIP-score, we directly compute the similarity between each image and the text, then average the scores across all views under the given setting to obtain the final score. Details of the prompt library are provided in \cref{sec:suppl_prompt}.

\subsection{Additional Qualitative Results}
\label{sec:suppl_qualitative}

\begin{figure*}[ht]
    \centering
    \includegraphics[width=.95\linewidth,trim=0 8em 0 5em,clip]{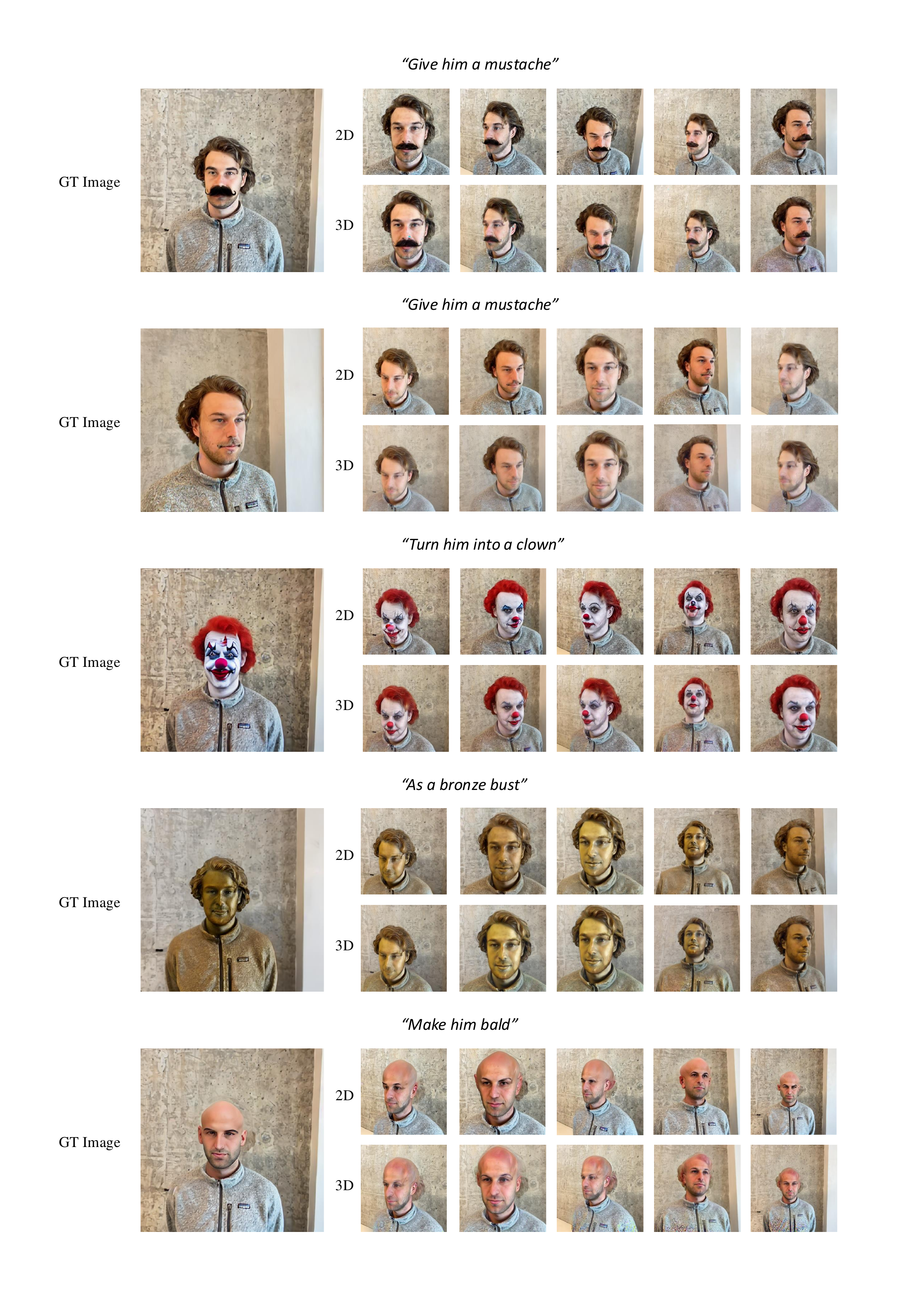}
    \caption{More qualitative results of our method C$^3$Editor.}
    \label{fig:suppl_gt_qual}
\end{figure*}

\begin{figure*}[ht]
    \centering
    \includegraphics[width=.95\linewidth,trim=0 8em 0 5em,clip]{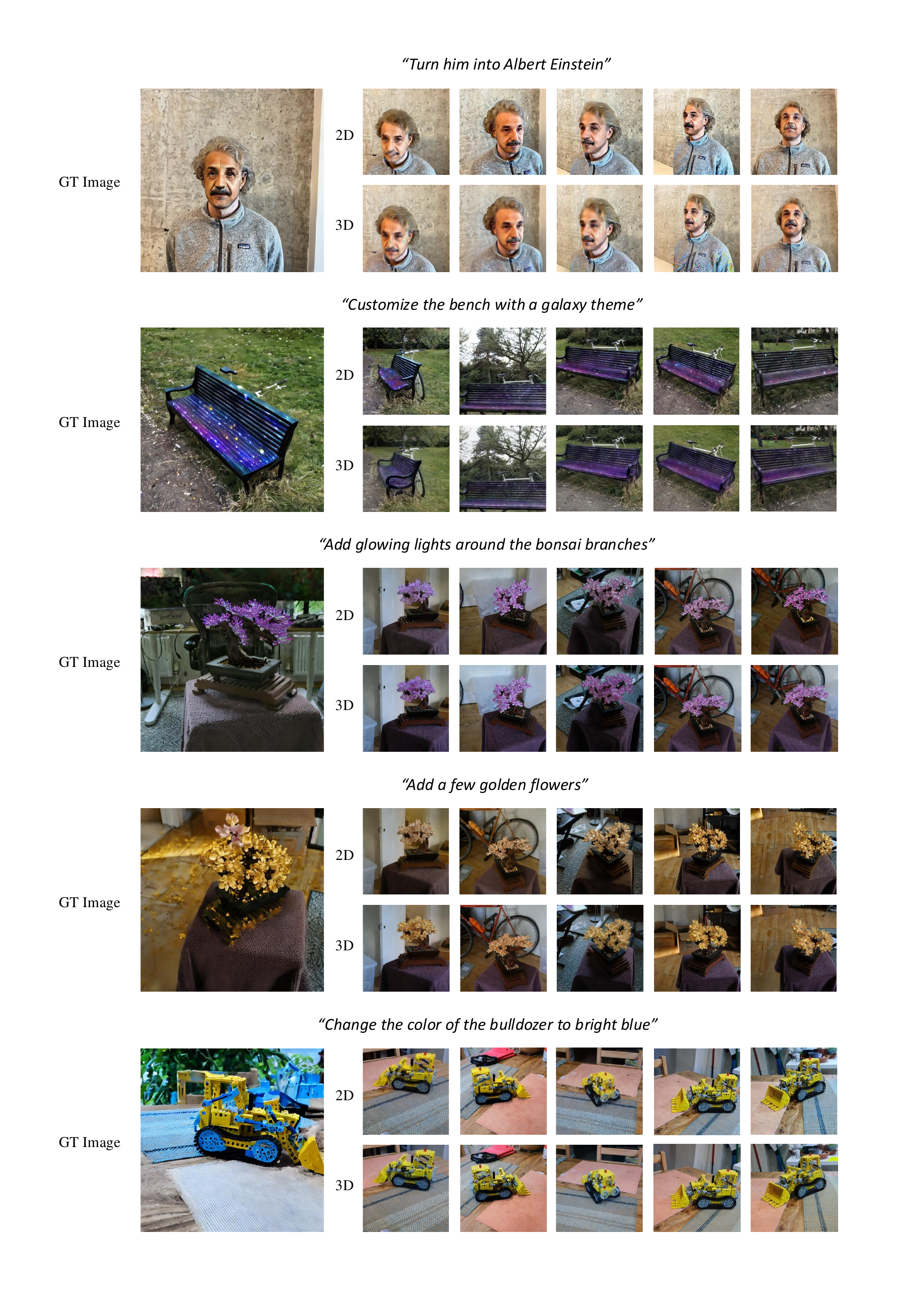}
    \caption{More qualitative results of our method C$^3$Editor.}
    \label{fig:suppl_gt_qual_2}
\end{figure*}

We provide additional qualitative results of our method C$^3$Editor on the MipNeRF-360~\cite{barron2022mip} and Instruct-NeRF-to-NeRF~\cite{haque2023instruct} datasets. As shown in \cref{fig:suppl_gt_qual} and \cref{fig:suppl_gt_qual_2}, our method successfully achieves controllable consistency in 2D models for 3D editing. The results demonstrate that our method can generate high-quality 3D editing results with controllable consistency across multiple views. The edited results are consistent with the provided editing text, the chosen GT edited image, and maintain high-quality image generation across different views. The results also show that our method can handle diverse editing scenarios, such as changing the color of objects, adding new objects, and modifying object shapes. The qualitative results further demonstrate the effectiveness of our method in achieving controllable consistency in 2D models for 3D editing.

\subsection{Ablation of View Propagation}
\label{sec:suppl_ablation}

We demonstrate the importance of view propagation in our method C$^3$Editor. As shown in \cref{fig:suppl_gt_view_prop}, we show the results of our method without view propagation. Without view propagation, the edited results exhibit inconsistencies across different views; only views near the GT view are consistent with the GT edited image. In contrast, views far from the GT view exhibit significant inconsistencies with the GT edited image. This may be caused by the limited generalization ability of the 2D editing model. Our method with view propagation can effectively address this issue by propagating the editing information across different views, achieving controllable consistency in 2D models for 3D editing, as shown in Fig.6 in \textit{Main Paper}.

\subsection{Visualization of GT Fitting}
\label{sec:suppl_gt}

We provide the visualization of ground truth fitting in our method C$^3$Editor. As shown in \cref{fig:suppl_fitting}, we visualize the 2D editing results of the GT view during intra-GT prior fitting in Sec. 3.3 in \textit{Main Paper}. The results demonstrate that our method can effectively fit the ground truth edited image, achieving controllable consistency in 2D models for 3D editing. The results further demonstrate the effectiveness of our method in achieving controllable consistency in 2D models for 3D editing.

\subsection{Prompt Library}
\label{sec:suppl_prompt}

``Give him a cowboy hat", ``Give him a mustache", ``Make him bald" ``Turn him into a clown", ``As a bronze bust", ``Turn him into Albert Einstein", ``Turn his face into a skull", ``Turn him into a Modigliani painting", ``Turn him into Batman", ``Turn him into Hulk", ``Turn him into an old lady", ``make it snowy", ``make the bike on fire", ``make the bench on fire", ``make the road snowy", ``make bike colorful", ``make the bike red and blue", ``make the bench red and blue", ``make the front wheel red and the rare wheel blue", ``Customize the bench with a galaxy theme", ``Add fallen autumn leaves", ``turn it into a marble table", ``Transform the table surface to white ceramic with blue patterns", ``Apply a gradient effect on the table", ``make it look like Van Gogh's painting", ``Add glowing lights around the bonsai branches", ``Add a few golden flowers", ``Make it look like it's covered in snow", ``Replace the flowers with glowing lanterns", ``Turn the bonsai flowers into red maple leaves", ``Change the color of the bulldozer to bright blue"

\end{document}